# Spin-Hall effect and emergent antiferromagnetic phase transition in n-Si

Paul C Lou[1], and Sandeep Kumar[1,2*]

[1] Department of Mechanical Engineering, University of California, Riverside, CA

[2] Materials Science and Engineering Program, University of California, Riverside, CA

**Abstract**

Spin current experiences minimal dephasing and scattering in Si due to small spin-orbit coupling and spin-lattice interactions is the primary source of spin relaxation. We hypothesize that if the specimen dimension is of the same order as the spin diffusion length then spin polarization will lead to non-equilibrium spin accumulation and emergent phase transition. In n-Si, spin diffusion length has been reported up to 6 μm. The spin accumulation in Si will modify the thermal transport behavior of Si, which can be detected with thermal characterization. In this study, we report observation of spin-Hall effect and emergent antiferromagnetic phase transition behavior using magneto-electro-thermal transport characterization. The freestanding Pd (1 nm)/ $Ni_{80}Fe_{20}$ (75 nm)/ MgO (1 nm)/ n-Si (2 μm) thin film specimen exhibits a magnetic field dependent thermal transport and spin-Hall magnetoresistance behavior attributed to Rashba effect. An emergent phase transition is discovered using self-heating 3ω method, which shows a diverging behavior at 270 K as a function of temperature similar to a second order phase transition. We propose that spin-Hall effect leads to the spin accumulation and resulting emergent antiferromagnetic phase transition. We propose that the length scale for Rashba effect can be equal to the spin diffusion length and two-dimensional electron gas is not essential for it. The emergent antiferromagnetic phase transition is attributed to the site inversion asymmetry in diamond cubic Si lattice.

Key words- Silicon, spin-Hall magnetoresistance, antiferromagnetic phase transition, Rashba effect,

**Introduction**

Due to small intrinsic spin orbit coupling (SOC), spin-phonon interaction is the primary spin relaxation mechanism in Si. We hypothesize if the spin diffusion length is larger than the specimen dimension then spin-phonon relaxation will be suppressed and non-equilibrium spin accumulation will occur as shown in Figure 1 a. In case of Si, the spin accumulation will lead to change in phonon in thermal transport due to spin-phonon relaxation behavior. The spin diffusion length for most of the materials is less than 10 nm but that of n-Si has been measured to be up to ~6 μm[1-4]. The long spin diffusion length allows us to observe the effect of spin polarization on phononic thermal transport in n-Si. The electrical spin injection and thermal spin-Seebeck tunneling are the popular methods of spin injection in Si. We hypothesized that electrical current across the ferromagnet/n-Si bilayer may lead to spin polarization in n-Si layer either due to spin-Hall effect (SHE) or the spin-Seebeck tunneling due to out of plane temperature gradient. When an electrical bias is applied across the conducting thin film specimen, a parabolic temperature gradient develops across the length of the specimen. In addition, out of plane temperature gradient may occur, which may lead to spin-Seebeck tunneling. The longitudinal temperature gradient gives rise to thermal transport across the specimen and the in-plane temperature gradient can be used to characterize the thermal properties (thermal conductivity and heat capacity). The spin polarization due to SHE or spin-Seebeck tunneling in n-Si specimen will modify the thermal transport behavior. The resulting change in thermal transport can be discovered using thermal property characterization also known as self-heating 3ω method[5-8].

**Experimental setup**

The self-heating 3ω method relies on the solution of the one-dimensional heat conduction equation for the specimen, which is given by

$$\rho C_p \frac{\partial \theta(x,t)}{\partial t} = \kappa \frac{\partial^2 \theta(x,t)}{\partial x^2} + \frac{I_o^2 \sin^2 \omega t}{LS}(R_o + R'\theta(x,t)), \quad (1)$$

where L and S are the length between the voltage contacts and the cross-sectional area of the specimen, respectively. $\rho$, $C_p$ and $\kappa$ are the density, specific heat and thermal conductivity in the material. $R_0$ is the initial electrical resistance of the specimen at temperature $T_o$. $R'$ is the temperature derivative of the resistance $R' = \left(\frac{dR}{dT}\right)_{T_o}$ at $T_o$. $\theta(x,t) = T(x,t) - T_o$ is the temporal ($t$) and spatial ($x$) dependent temperature change, as measured along the length of the specimen, which coincides with the heat flow direction. The $V_{3\omega}$ response is approximately

$$V_{3\omega} \approx \frac{4I^3 R_o R' L}{\pi^4 S \kappa \sqrt{1+(2\omega\gamma)^2}} \quad (2)$$

where $\gamma$ is the thermal time constant and is related with the heat capacity $\left(C_p = \frac{\pi^2 \gamma \kappa}{\rho L^2}\right)$. The $V_{3\omega}$ is a function of both thermal conductivity and heat capacity. The thermal conductivity can be expressed in terms of the third harmonic voltage $V_{3\omega}$ in the low frequency limit ($\omega\gamma \to 0$) by

$$\kappa \approx \frac{4I^3 R_o R' L}{\pi^4 V_{3\omega} S} \quad (3)$$

We can infer that the heat capacity and thermal conductivity can be considered as a function of resistance and $V_{3\omega}$ response ($f(\kappa, C_p) = \frac{R}{V_{3\omega}}$). The self-heating 3ω method has been successfully applied to elucidate the spin mediated thermal transport behavior in p-Si specimen[9, 10]. The self-heating 3ω method requires a freestanding thin film specimen to minimize the heat loss and in turn error in thermal property measurement.

To fabricate the freestanding thin-film structure, we started with a commercially available silicon on insulator (SOI) wafer with a P-doped 2 μm thick device layer having a resistivity of 0.001-0.002 Ω cm. Using UV photolithography, we patterned and etched the front setup in the Si device layer using deep reactive ion etching (DRIE). A freestanding Si structure is made by etching the buried oxide using hydrofluoric acid vapor etching. In the next step, we removed the surface oxide by Ar milling for 10 minutes, followed by deposition of 1 nm MgO using RF sputtering. A layer of 25 nm $Ni_{80}Fe_{20}$/1 nm Pd is then deposited onto the device using e-beam evaporation. The material deposition using evaporation leads to line-of-sight thin film deposition on the top of n-Si layer only. The MgO layer acts as a tunneling barrier for spin transport across the interface between the Si and the $Ni_{80}Fe_{20}$/Pd layers. The experimental setup with a freestanding multilayer specimen (*l*-170 μm, *w*-9 μm and *t*-2 μm) consisting of Pd (1 nm)/ $Ni_{80}Fe_{20}$ (25 nm)/MgO (1 nm)/n-Si (2 μm) is shown in Figure 1 b.

**Results and discussion**

For self-heating 3ω method, we apply an alternating current (ac) bias across the four-probe specimen. We acquire the $V_{1\omega}$ (electrical resistance), $V_{2\omega}$ (spin mediated thermoelectric effects including spin-Seebeck effect (SSE)[11, 12] and anomalous Nernst effect (ANE)[12]) and $V_{3\omega}$ (thermal properties)[9, 10] responses as a function of temperature and magnetic field. The magneto-electro-thermal transport measurements are carried out using a Quantum Design physical property measurement system (PPMS) at high vacuum. The self-heating 3ω method requires a cubic relationship between heating current and the corresponding $V_{3\omega}$ response. We observe that the device having 25 nm of $Ni_{80}Fe_{20}$ layer deviates significantly from the cubic relationship. We assumed that the resistance of 25 nm of $Ni_{80}Fe_{20}$ layer will be larger than the 2 μm of n-Si layer, which will keep the specimen resistance behavior Ohmic. We observed that the n-Si layer is significantly more conducting than the $Ni_{80}Fe_{20}$ layer. Although the specimen design is not expected to be the source of error in cubic relationship but we fabricated the new set of devices with 75 nm $Ni_{80}Fe_{20}$ layer to maintain equality of current density within each layer. However, the new device does not exhibit the cubic relationship, between current and the $V_{3\omega}$ response, as well.

To uncover the reason for this deviation, we measured the $V_{1\omega}$, $V_{2\omega}$ and $V_{3\omega}$ responses as a function of temperature at 0.3 K/min from 300 K to 5 K at applied heating current of 1.55 mA at 7 Hz on the 75 nm $Ni_{80}Fe_{20}$ device. For self-heating 3ω method, the metallic behavior is essential and the $R_{1\omega}$ shows an Ohmic behavior as shown in Figure 2 a. We analyze the $V_{3\omega}$ and $\frac{R}{V_{3\omega}}$ responses, which show an inflection and diverging behavior respectively at 270 K as shown in Figure 2 c and d. The $Ni_{80}Fe_{20}$ has a thermal conductivity of ~20 W/m·K [13], which is significantly lower than that of Si (~80 W/m·K[14]). Hence the observed thermal transport is attributed primarily to the Si layer only. For in-plane conduction, we estimated that the 2 μm n-Si is 2000 times $\left(R_{thermal} = \frac{l}{\kappa A}\right)$ more thermally conducting than the 75 nm $Ni_{80}Fe_{20}$ layer. We attributed the deviation from cubic relationship between current and $V_{3\omega}$ response due to this emergent behavior. Since the emergent behavior can only be observed in thermal transport measurement ($V_{3\omega}$ and $\frac{R}{V_{3\omega}}$ responses) and not in resistance ($R_{1\omega}$) measurement, it originates in n-Si layer. Around the transition phase, equation 2 is not a good approximation to thermal conductivity and equation 1, which also include heat capacity, will be a better approximation of thermal transport behavior. The diverging behavior in $\frac{R}{V_{3\omega}}$ can be considered a second order phase transformation since the second order phase transformations are characterized from singularities or discontinuities in the temperature dependent heat capacity measurements [15-21]. The ratio of $\frac{R}{V_{3\omega}}$ experiences large increase again below 50 K, indicating another gradual phase transition. To verify if behavior is intrinsic to n-Si, measurements are repeated on a control n-Si specimen with 0.55 mA$_{rms}$ heating current, as shown in Figure 2. The control specimen exhibits the Ohmic behavior as expected for a highly doped n-Si, but it does not show transition in $V_{3\omega}$ and $\frac{R}{V_{3\omega}}$ responses. The temperature dependent $\frac{R}{V_{3\omega}}$ response, shown in Figure 2 d, is similar to the thermal conductivity of highly doped n-Si reported in the literature. The comparison of two results indicated the transition is present in the specimen with a ferromagnetic layer. We propose that spin accumulation in n-Si is the underlying cause of the observe behavior. The spin accumulation leads to emergent antiferromagnetic phase transition due to site inversion asymmetry in Si. To uncover the mechanism for the transition, we undertake magnetic field dependent measurements.

The magnetic field (2 T to -2 T) dependent $V_{1\omega}$, $V_{2\omega}$ and $V_{3\omega}$ responses are acquired at 300, 270, 225, 175, 100, 50 and 5 K, as shown in Figure 3. The temperatures were chosen around the valley and plateau of the transition temperatures seen in $V_{3\omega}$ and $\frac{R}{V_{3\omega}}$ responses in Figure 2 c-d. We estimate the resistance of 75 nm Ni$_{80}$Fe$_{20}$ layer to be ~84 Ω and the n-Si layer to ~190 Ω. The observed magnetoresistance behavior pertains to the Ni$_{80}$Fe$_{20}$ layer. The magnetoresistance increases as the temperature is reduced from 300 K to 5 K. The negative magnetoresistance increases from 1.1% at 300 K to 3.375% at 5 K. Below 200 K (after the emergent transition), the observed magnetoresistance shows a knee at ~1.25 T, which corresponds to the saturation magnetization of Ni$_{80}$Fe$_{20}$ layer as shown in Figure 3 a. At high temperatures of 300 and 270 K, specimen behave like a soft magnetic, with saturation field of 0.3 T—four times less than the expected saturation field of Ni$_{80}$Fe$_{20}$. Since Ni$_{80}$Fe$_{20}$ is the only ferromagnetic material in the specimen, the reduction of saturation field indicates spin-orbit torque (SOT) transfer from n-Si to Ni$_{80}$Fe$_{20}$ layer. The expected saturation field of 1.25 T in Ni$_{80}$Fe$_{20}$ is seen at low temperatures of

100, 50 and 5 K, indicating absence of SOT. In $V_{2\omega}$ sweep, effect of $Ni_{80}Fe_{20}$ switching is observed, as shown in Figure (b). At 300 and 100 K, $V_{2\omega}$ switching behavior is observed after crossing the zero-magnetic field, as expected for switching of ferromagnetic material. However, at 50 K and below, $V_{2\omega}$ switching occurs prior to zero magnetic field. The inverted switching behavior is attributed to the antiferromagnetic canted spin states in n-Si due to spin accumulation, which has been reported in p-Si at low temperatures[10]. The measured $V_{3\omega}$ dependency with field is shown in Figure 3c. At 300 K, $V_{3\omega}$ response increases with increasing field. But as the temperature is reduced below 270 K, the relationship inverts; the $V_{3\omega}$ response decreases with increasing field. Inversion of relationship at 270 K between external field and thermal transport behavior supports our assertion of emergent phase transition behavior. Maxima of $V_{3\omega}$ is seen at 170 K, agreeing with the maxima in temperature sweep from Figure 2 c. Saturation of $Ni_{80}Fe_{20}$ layer became suddenly profound at temperatures below 270 K, but is only gradually visible in $R_{1\omega}$. The difference with electrical and thermal transport is seen by comparing field response between $V_{1\omega}$ and $V_{3\omega}$. Since both $Ni_{80}Fe_{20}$ and n-Si layers are metallic, $R_{1\omega}$ indicate the behavior of electrical transport within both layers. Whereas n-Si is the primary heat carrier in the specimen with dominant phonon population, majority of the effects seen in $V_{3\omega}$ is the contribution from thermal properties of n-Si. Thus, the phase transition seen in $V_{3\omega}$ response is attributed to the n-Si layer as stated earlier. Since the transition existed only with presence of ferromagnetic layer on top of n-Si, $Ni_{80}Fe_{20}$ layer induces spin accumulation in n-Si. This leads to the strong spin-phonon relaxation behavior in n-Si, which changes the phonon mediated thermal transport (since phonon is the primary spin relaxation mechanism in Si).

We attribute the spin polarization to SHE in n-Si. To confirm the presence of spin polarization within n-Si, the behavior of the $V_{1\omega}$, $V_{2\omega}$ and $V_{3\omega}$ responses are acquired as a function of angular rotation of magnetic field in the yz-plane. If the magnetization direction of ferromagnet is orthogonal to the spin at the interface, spin would be absorbed. If the directions were parallel instead, spin would be reflected. The reflected spin current would be converted back into charge current through inverse spin-Hall effect (ISHE). The reflection of spin current due to SHE from the ferromagnetic interface gives rise to spin Hall magnetoresistance (SMR) behavior[22-26]. The measurement is performed in the zy-plane at 8 T at 300, 200, 100, 50 and 5 K, as shown in Figure 4. At 300 and 200 K, the specimen exhibits SMR behavior, with decreasing magnitude as temperature is decreased. For temperatures below 100 K, magnetoresistance exhibits anisotropic magnetoresistance (AMR) behavior. The presence of SMR at high temperatures and AMR at low temperatures indicates both signals occur simultaneously, but the dominant phenomena for the overall signal is dependent on the specimen temperature. The origin of SMR must be from n-Si since SMR behavior can not originate from spin tunneling (spin-Seebeck tunneling or electrical injection) from $Ni_{80}Fe_{20}$ layer. In addition, n-Si has insignificant intrinsic spin-orbit coupling. Hence, inverse spin-Hall effect(ISHE) and SMR behavior arising intrinsically in n-Si should not be observable. We propose that the interfacial spin-orbit coupling gives rise to ISHE and SMR observed in this study. We do not observe a sinusoidal behavior expected for SSE and ANE from the $V_{2\omega}$ response shown in Figure 4 b. We observe a dominant $\sin^2\theta_{zy}$ behavior in the $V_{2\omega}$ response, which can be attributed to the spin-phonon interactions[9] due to SHE. In addition, the observed $V_{2\omega}$ response can arise from the Rashba effect mediated tunneling anisotropic thermopower[27]. While the transition from SMR to AMR occurs below 200 K, we observe the emergent phase transition $V_{3\omega}$ response below 300 K similar to temperature dependent measurement. At 300 K, we observe a $\sin^2\theta_{zy}$ behavior in the $V_{3\omega}$ response with negative amplitude. The magnitude shows a sign reversal as the temperature is lowered to 200 K. The amplitude increases with decrease in temperature as shown in Figure 4 c. The observed behavior is attributed to the SHE mediated thermal resistance and hence called as spin-Hall magneto thermal resistance (SMTR) [9].

From these measurements, we propose that the SHE in n-Si causes non-equilibrium spin accumulation due to proximity with $Ni_{80}Fe_{20}$ layer. The spin accumulation leads to the observed emergent antiferromagnetic phase transition behavior as hypothesized (Figure 1 a). Si is diamagnetic in the bulk form but local (weak) antiferromagnetic interactions have been predicted to exist due to site inversion asymmetry of centosymmetric diamond cubic lattice. The emergent antiferromagnetic phase transition is attributed to the site-inversion asymmetry in inversion symmetric diamond cubic lattice of n-Si[28, 29]. The emergent antiferromagnetic phase transition changes the magneto-thermal transport behavior as shown in Figure 2 c-d. The applied magnetic field causes dephasing of spin excitations in emergent antiferromagnetic phase and changes spin-phonon relaxation behavior. This enhances the thermal transport behavior at low temperatures, as observed in reduction in $V_{3\omega}$ response. The observation of SHE is supported by the SMR measurement at 300 K and 200 K. This is the first experimental proof of SHE in n-Si. The SMR behavior disappears at low temperatures due to emergent antiferromagnetic phase transition. But, n-Si does not have intrinsic spin-orbit coupling. We propose that the ISHE, which is essential for SMR behavior, occurs at or near the interface of $Ni_{80}Fe_{20}$/MgO/n-Si. This interface gives rise to structure inversion asymmetry and in turn Rashba spin-orbit coupling[30-32]. The spin-orbit coupling due to structure inversion asymmetry in Si metal-oxide semiconductor field effect transistor (MOSFET) has been reported in magneto-transport behavior in two-dimensional electron gas (2DEG) in at low carrier concentrations [33-36]. In addition, spin resonance measurements on Si metal-oxide semiconductor field effect transistor (FET) report suppression of spin resonance due to SOC[37]. This behavior agrees with the proposed hypothesis presented in this study due to ferromagnetic metal-oxide-Si interface except we observe this behavior at higher charge carrier concentrations. For strong Rashba SOC, the essential requirements are structure inversion asymmetric interface and intrinsic SOC of layer materials. In the Si MOSFET, the SOC due to structure inversion asymmetry is relatively small because the gate metals have small intrinsic SOC. In this study, n-Si have insignificant intrinsic SOC but $Ni_{80}Fe_{20}$ have significantly large intrinsic SOC[38], which may give rise to the strong Rashba SOC due to proximity effect [3, 39]. This poses a problem since Rashba effect is expected at two-dimensional electron gas (2DEG) or nanoscale thin films (few nanometer) whereas n-Si thin in this study is 2 μm. The length scale for Rashba effect is currently unknown. We propose that the length scale for Rashba spin-orbit coupling can be as large as the spin diffusion length in semiconductor or normal metal, which is supported by the observed experimental results. The observed Rashba effect may challenge the non-local spin transport measurement[1, 3, 40] in Si since Rashba effect may enhance the spin polarization.

**Conclusion**

In conclusion, we report magneto-electro-thermal transport measurements in specimen composed of Pd (1 nm)/$Ni_{80}Fe_{20}$ (75 nm)/MgO (1 nm)/n-Si (2 μm). The temperature-dependent measurement of $R_{1\omega}$, $V_{2\omega}$ and $V_{3\omega}$ indicates a spin mediated emergent phase transition at 270 K. The emergent phase transition is uncovered from the diverging behavior in $R_{1\omega}/V_{3\omega}$, which is related to the thermal transport/properties. The magnetic field-dependent $V_{3\omega}$ response support the transition behavior. In addition, the magnetoresistance behavior reveals existence of SOT transfer from n-Si to $Ni_{80}Fe_{20}$ layer, which diminished as temperature decreased. The angular rotation of magnetic field in yz-plane shows existence of SMR behavior at 300 and 200 K, and AMR is observed at and below 170 K. The AMR originates from $Ni_{80}Fe_{20}$ layer and SMR originates from the SHE in n-Si coupled with interfacial ISHE due to Rashba effect. With n-Si having long spin diffusion length, we propose that SHE leads to non-equilibrium spin accumulation. The spin accumulation leads to an antiferromagnetic emergent phase transition. The emergent behavior can only be observed in thermal response behavior due to spin-phonon interactions in n-Si and phonons being the primary heat carrier in n-Si. The observation of magneto-thermal transport behavior, emergent phase transition and interfacial Rashba spin-orbit coupling may lead to realization of Si spintronics.

List of Figures

Figure 1. (a) The hypothesis of the emergent antiferromagnetic phase transition and (b) the false color scanning electron micrograph showing the experimental setup.

Figure 2. The temperature-dependent measurements from 300 to 5 K for Pd (1 nm)/ $Ni_{80}Fe_{20}$ (25 nm)/MgO (1 nm)/n-Si (2 μm) and control n-Si device for (a) Electrical resistance $R_{1\omega}$ (b) $V_{2\omega}$ response (c) the $V_{3\omega}$ response, and (d) The $\frac{R}{V_{3\omega}}$ response (function of thermal conductivity and heat capacity).

Figure 3. The magnetic field dependent measurements for an applied magnetic field of 2T- -2 T showing (a) magnetoresistance, (b) the change in $V_{2\omega}$ response and (c) the change in $V_{3\omega}$ response at 300 K, 270 K, 225 K, 175 K, 100 K, 50 K and 5 K.

Figure 4. The angular dependence of magnetoresistance for an applied magnetic field rotated in the yz-plane showing SMR behavior at 300 K and 200 K and AMR behavior at 100 K, 50 K and 5 K.

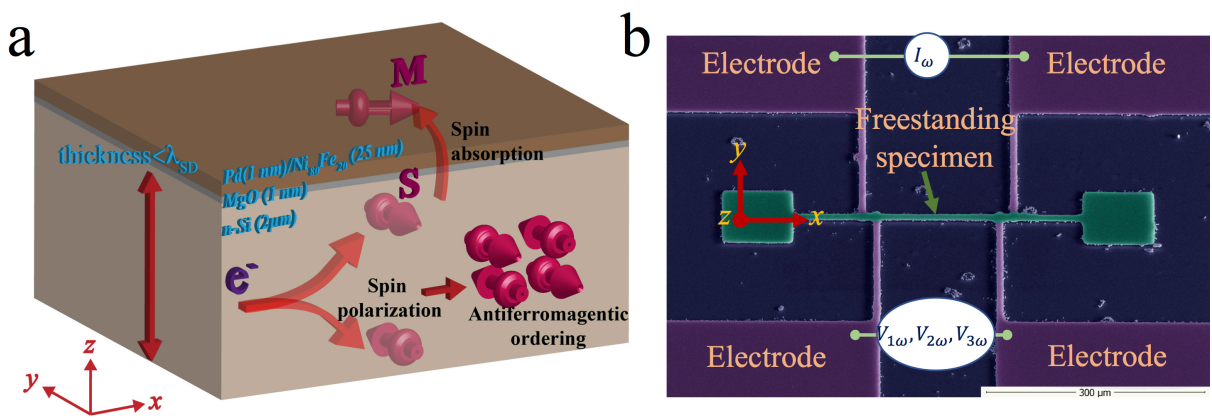

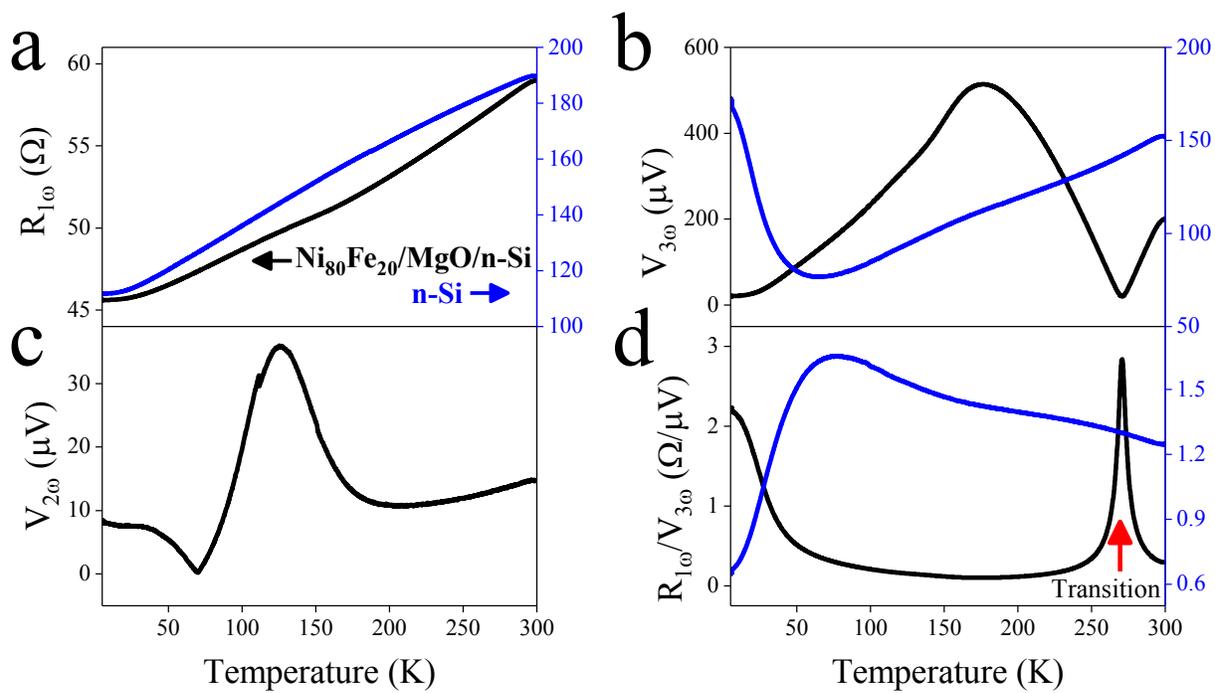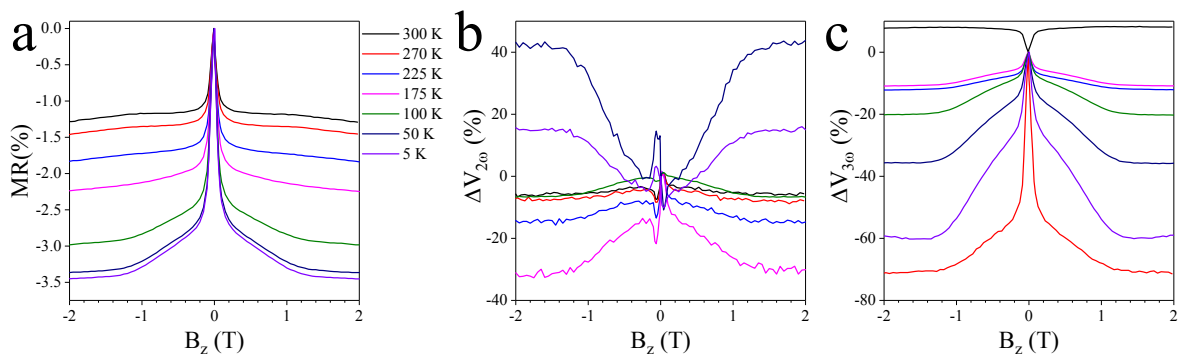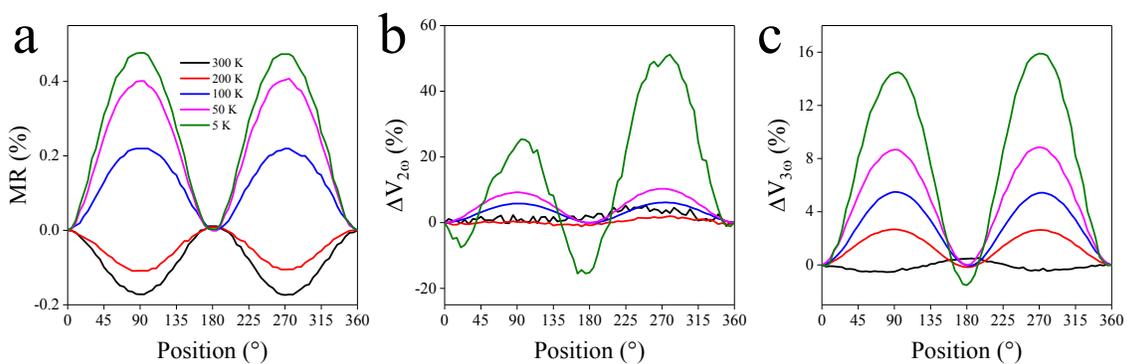